\documentclass[12pt]{article}
\usepackage{amsmath,amssymb,amsfonts,amsthm}
\title{New nonlinear coherent states associated to inverse bosonic and $f$-deformed
       ladder operators}
\author{M. K. Tavassoly
\\
\footnotesize{Atomic and Molecular Group, Faculty  of Physics, Yazd University, Yazd, Iran}
\\ \footnotesize{e-mail: mktavassoly@yazduni.ac.ir  } }

\begin{document}


\newcommand{\I}{\mathbb{I}}
\newcommand{\norm}[1]{\left\Vert#1\right\Vert}
\newcommand{\abs}[1]{\left\vert#1\right\vert}
\newcommand{\set}[1]{\left\{#1\right\}}
\newcommand{\R}{\mathbb R}
\newcommand{\C}{\mathbb C}
\newcommand{\DD}{\mathbb D}
\newcommand{\eps}{\varepsilon}
\newcommand{\To}{\longrightarrow}
\newcommand{\BX}{\mathbf{B}(X)}
\newcommand{\HH}{\mathfrak{H}}
\newcommand{\D}{\mathcal{D}}
\newcommand{\N}{\mathcal{N}}
\newcommand{\W}{\mathcal{W}}
\newcommand{\RR}{\mathcal{R}}
\newcommand{\HD}{\hat{\mathcal{H}}}
 \maketitle

\begin{abstract}
  Using the {\it nonlinear coherent states method},
  a formalism for the construction of the coherent states associated
  to {\it "inverse bosonic operators"} and their  dual family
  has been proposed.  Generalizing the approach, the
  "inverse of $f$-deformed ladder operators" corresponding to the
  nonlinear coherent states in the  context of
  quantum optics and the associated  coherent states have been
  introduced.  Finally, after applying the proposal to
  a few known physical systems, particular nonclassical features as
  sub-Poissonian statistics and the squeezing
  of the quadratures of the radiation field corresponding to the
  introduced states have been investigated.
 \end{abstract}



  \section{Introduction}\label{sec-intro}

    The {\it standard coherent states} $|z\rangle $ may be obtained from the action of
    displacement operator on the vacuum [1],
\begin{equation}\label{displace}
   D(z)=\exp(  z a^\dag - z^* a),
   \qquad D(z) | 0\rangle=|z  \rangle,
\end{equation}
   or the right eigenstate annihilation operator
\begin{equation}\label{eigen-css}
   a|z\rangle = z |z\rangle,
\end{equation}
   where $z\in \C$ and  $a$, $a^\dag$ are the standard bosonic annihilation, creation
   operators, respectively.  The states $| z \rangle$ are also
   minimum uncertainty states. It is well known that the expansion of these states in the
   Fock space is as follows:
\begin{equation}\label{SCS}
  |z\rangle = \exp ^{-\frac 1 2 |z|^2}\sum_{n=0}^\infty
  \frac {z^n}{\sqrt {n!}}\;| n  \rangle,
\end{equation}
   where the set $\left\{|n\rangle\right\}_{n=0}^\infty$
   is the number states
   of the quantized harmonic  oscillator with hamiltonian $\hat H=a^\dag a + \frac 1 2$.


  Due to the fact that the bosonic annihilation and creation
  operators, $a$ and $a^\dag$ are singular operators,
  the inverse operators $a^{-1}$ and ${a^\dag}^
  {-1}$ are not well-defined.
  Nevertheless, the following generalized operators may be found in
  literature through the actions  [2, 3]
\begin{equation}\label{inv-crea-op}
  {a^{\dag}}^{-1}|n\rangle = (1-\delta_{n,0})
  \frac{1}{\sqrt{n}}|n-1\rangle,
\end{equation}
   and
\begin{equation}\label{inv-ann-op}
  a^{-1}|n\rangle = \frac{1}{\sqrt{n+1}}|n+1\rangle,
\end{equation}
  where $\delta_{n,0}$ is $1$ when $n=0$, otherwise it is $0$
  (so by definition ${a^{\dag}}^{-1}|0\rangle =0$).
  Obviously, $a^{-1}$ $({a^{\dag}}^{-1})$ behaves like creation (annihilation) operator.
   Also, the statement that $a^{-1}$
  (${a^\dag}^{-1}$) is the right (left) inverse of $a$
  ($a^\dag$) seems to be legally true, since:
  \begin{eqnarray}\label{inv-r-l}
  a a^{-1} &=& {a^\dag}^{-1} a^\dag =\hat I\; ,\nonumber\\
  a^{-1} a &=& a^\dag {a^\dag}^{-1}  =\hat I -|0\rangle\langle0| \:,
  \end{eqnarray}
  where $\hat I$ is the unit operator.
  Thus one may get the commutation relations
  $[a, a^{-1}] = |0 \rangle \langle 0 | =[{a^\dag}^{-1},
  a^\dag]$.
  Anyway, the usefulness of these inverse operators in various contexts can be
  found in previous publications we shall address at this point.
  Indeed, the operators in
  (\ref{inv-crea-op}) and (\ref{inv-ann-op}) have been studied in
  para-bose particles. In addition, these operators enable one to
  find the eigenvalue equation for the squeezed coherent states.
  The vacuum and the first exited squeezed states are the eigenstates of
  ${a^{\dag}}^{-1} a$ and $ a {a^{\dag}}^{-1}$,  respectively [4].
   Therefore, for instance, since the squeezed vacuum states
   can be generated through some nonlinear optical processes,
   it is remarkable that  inverse bosonic operators may
   play an important role in studying the time evolution of some nonlinear systems.
   Also, the important role of these operators has been followed in
   metaplectic group structure of $Mp(2)$ group, which
   is a two-fold cover of $Sp(2,\; R)$ and $SU(1,\; 1)$ groups  [5].
   Let us recall that {\it "photon added coherent states"} first
   introduced by Agarwal and Tara [6] are
   closely related to the inverse bosonic operators
   [3], where it has been shown that these states denoted by $|z, m\rangle=
   {a^\dag}^m|z\rangle$ are eigenstates of the operator $a - m
   {a^\dag}^{-1}$ with eigenvalues $z$. Subsequently, {\it " photon subtracted
   coherent states"} can be obtained by $m$-times actions of
   $a^{-1}$ on $|z\rangle$,  followed by $m$-times actions of ${a^\dag}^{-1}$
   on the resultant states, i.e. $|z, -m\rangle = {a^\dag}^{-m} a^{-m} |z\rangle$.

    Nowadays generalization of coherent states  besides their experimental generations
    have made much interest in quantum physics, especially in quantum optics [1].
    These  quantum states exhibit some interesting  {\it "non-classical
    properties"} particularly squeezing, antibunching, sub-Poissonian
    statistics and oscillatory number distribution.
    Along achieving this goal, the first purpose of the present paper is to outline a
    formalism for the construction of coherent states associated
    to the {\it "inverse bosonic operators"},
    the states that have not been found in the literature up to know.
    But before paying attention to the this matter,
    a question may naturally arise about the relation between
    the operators ${a^\dag}^{-1}$ and $a^{-1}$ and
    the standard coherent states
    $|z\rangle$.  Due to the second equation in
    (\ref{inv-r-l}) it
    is readily found that the following does not hold: $a^{-1} |z\rangle = z^{-1}
    |z\rangle$, which at first glance may be expected from equation (\ref{eigen-css}).
    Instead,
    one has
\begin{equation}
    a^{-1}|z\rangle = z^{-1}\left[|z\rangle
    -e^{-\frac{1}{2}|z|^2}|0\rangle  \right].
\end{equation}
   This is consistent with the fact that right eigenstate of the
   operator $a^{-1}$ does not exist,  originates from the creation-like
   characteristic of  $a^{-1}$ [2].
   Also it can be seen that the standard coherent state $|z\rangle$
   in (\ref{SCS}) is not the eigenstate of ${a^\dag}^{-1}$, one has instead
\begin{equation}\label{}
   {a^\dag}^{-1}|z\rangle = z e^{-\frac 1 2 |z|^2}
   \sum_{n=0}^\infty \frac{z^n}{\sqrt {n!}(n+1)}
   |n\rangle\;.
\end {equation}

    Nevertheless, as we will observe, "nonlinear coherent state
    method" provides a rich enough
    mathematical structure allowing us to establish our aim.
    We will illustrate that although
    the presented formalism yields a nonnormalizable coherent state corresponding to
    the inverse bosonic operator (${a^\dag}^{-1}$),
    the associated dual family is well-defined.
    In the continuation of the paper, along
    generalization of the proposal to the
    {\it "inverse nonlinear ($f$-deformed) ladder operators"}
    involved in the nonlinear coherent states context in
    quantum optics, the associated  generalized coherent states have been also
    introduced.
    Then, as some physical realizations of the proposed formalism, hydrogen-like
    spectrum, harmonious states
    and Gilmore-Perelomov representation of $SU(1, 1)$ group have been considered.
    Taking into account their nonlinearity functions, we shall deduce the explicit
    form of the corresponding coherent states associated to the inverse $f$-deformed
    operators.
    At last, we conclude the paper with investigating some interesting nonclassical
    properties, for instance  sub-Poissonian statistics (anti-bunching) and the squeezing
    of the quadratures of the field of the obtained  states, numerically.

 \section{Coherent states of inverse bosonic operators}
         \label{sec-nl}

    In this section after presenting a brief review of the nonlinear coherent states,
    we are going to establish a link between the "inverse bosonic" and "$f$-deformed
     (nonlinear) ladder" operators.

 \subsection{The link between "inverse bosonic" and "$f$-deformed ladder" operators}
    The notion of {\it "nonlinear"} or {\it "$f$-deformed"}
    coherent states which provides  a powerful method to analyze a large number
    of the quantum optics states
    [7, 8, 9].
    Any class of these states, characterized by a particular intensity dependent
    function $f(n)$, is defined as the solution of the
    typical  eigenvalue  equation
    $a_f |z, f\rangle  = z |z, f\rangle $, with decomposition in the number states space as:
 \begin{equation}\label{NLCSf}
     |z, f\rangle = N(|z|^2)^{-\frac 1 2}\sum_{n=0}^\infty
     \frac {z^n}{\sqrt {n!}\;[f(n)]!}\;| n  \rangle,
 \end{equation}
     where  $a_f= a f(n)$ is the $f$-deformed annihilation operator,
     $[f(n)]! \doteq f(n)f(n-1)f(n-2)\dots f(2)f(1)$ and $[f(0)]! \doteq 1$.
     The function $N(|z|^2)$ in (\ref{NLCSf}) is the normalization
     constant can readily be calculated as $\sum_{n=0}^\infty |z|^{2n}/[nf^2(n)]!$.
     Choosing different $f(n)$'s lead to distinct generalized
     coherent states.

      The nonorthogonality (as a consequence of overcompleteness) of the states in (\ref{NLCSf}), i.e. $\langle z, f|z', f\rangle \neq 0 $ (and all the new states will be introduced in the present paper) is so clear matter
      we pay no attention to it.
      These states are required to satisfy  the resolution of the identity
   \begin{equation}\label{res-ffff}
      \int _D d^2 z |z, f\rangle W(|z|^2)\langle z, f | = \sum_{n=0}^\infty |n\rangle  \langle n|
      =\hat I
   \end{equation}
   where $d^2 z \doteq dx dy $,
   $W(|z|^2)$ is a positive weight function
   may be found after specifying $f(n)$, and $D$ is the domain of the states
   in the complex
   plane defined by the disk
  \begin{equation}\label{diskff}
    D = \{z \in \C, \; |z|\leq \lim_{n\rightarrow \infty}
    [nf^2(n)]\},
  \end{equation}
    centered at the origin in the complex plane.
    Inserting the explicit form of the states (\ref{NLCSf}) in (\ref{res-ffff}) with $|z|^2 \equiv x$
    it can be easily checked that the resolution of the identity
    holds if the following moment problem is satisfied:
  \begin{equation}\label{res-fff}
      \pi \int _0^R dx \sigma (x) x^n=
      [nf^2(n)]!,\qquad n=0,1,2,\cdots ,
   \end{equation}
    where $\sigma(x) = \frac{W(x)} {N(x)}$ and $R$ is the radius of convergence determined by the relation
    (\ref{diskff}). The condition (\ref{res-fff}) presents a severe restriction on the
    choice of $f(n)$. Altogether, there are cases for which the completeness of some previously
    introduced coherent states has been demonstrated  a few years later elsewhere (photon added coherent states
    introduced in 1991 [6] while their completeness condition demonstrated in 2001 [10]).
    In fact, only a relatively small number of  $f(n)$ functions are known, for which the functions $\sigma(x)$  can be
    extracted.

     The action of $f$-deformed
     creation operator defined as $a_f^\dag = f^\dag(\hat n) a^\dag $ on the number
     states expresses as follows
  \begin{equation}\label{af}
     a_f^\dag  |n\rangle = f^\dag(n+1)\sqrt {n+1} |n+1\rangle.
  \end{equation}
   Now, going back to our goal in the paper, comparing
   equations (\ref{af}) and (5) it can be easily seen that,
   \begin{equation}\label{f-bos}
  a^{-1} \equiv  a^\dag_f = f^\dag(\hat n) a ^\dag ,
  \quad \mbox{with}\quad f( \hat n)=\frac{1}{ \hat n}.
 \end{equation}
   Similarly, using the action of $f-$deformed annihilation operator $a_f$
   on the number states, i.e.
 \begin{equation}\label{}
    a_f|n\rangle= f(n) \sqrt n |n-1\rangle
 \end{equation}
  and then comparing with (\ref{inv-crea-op}) one readily finds
 \begin{equation}\label{}
  {a^\dag}^{-1} \equiv  a_f = a f (\hat n),
 \end{equation}
   with the same $f(\hat n)$ introduced in (\ref{f-bos}).
    Note that the following also holds
 \begin{equation}\label{}
   a^\dag f^\dag(\hat n)=f^\dag(\hat  n-1)a^\dag, \qquad f (\hat n) a = a f(\hat n-1).
 \end{equation}
   Therefore, taking into account all
   the above results we can write the explicit forms of the
   inverse bosonic  operators denoted by $a_f$ and $a^\dag_f$,
   and the related actions as follow [7]:
 \begin{equation}\label{af-non}
  a_f \equiv {a^\dag}^{-1} = a \; \frac {1} {\hat n},\qquad
  a_f |n\rangle = (1-\delta_{n,0})
  \frac{1}{\sqrt{n}}|n-1\rangle,
 \end{equation}
 \begin{equation}\label{af-1}
    a^\dag _f \equiv {a}^{-1} = \frac {1} {\hat n}\; a^\dag,  \qquad
    a^\dag _f |n\rangle = \frac{1}{\sqrt{n+1}}|n+1\rangle.
 \end{equation}
  Actually in the latter equations  the nonlinearity
  function is considered as $f(\hat n)=\frac{1}{\hat n}$. The
  equations (\ref{af-non}) and (\ref{af-1}) confirm us that $a_f$  ($a_f^\dag$)
  annihilates (creates) one (deformed) quanta of photon in some optical processes, respectively.
  For the commutation relation between the two ladder operators introduced in
  (\ref{af-non}) and (\ref{af-1}) one arrives at
 \begin{equation}\label{}
   [a_f,\; a^\dag_f] = - \frac {1} { \hat n (\hat n+1) }, \quad
   \mbox{for}\;\; n \neq 0,
   \end{equation}
 and
   \begin{equation}\label{}
    [a_f,\; {a^\dag}_f] = |0 \rangle \langle 0 |, \quad
    \mbox{for}\;\;
    n=0.
 \end{equation}
   Interestingly, this method with the factorized Hamiltonian formalism,
   permits one to derive a Hamiltonian responsible to the dynamics
   of the (inverse) system as [9]
 \begin{equation}\label{Hamilt-bob}
    \hat h  = a^\dag _f a_f = \frac {1} {\hat n} \equiv \hat H^{-1}
    \qquad \mbox{for}\;\; n \neq 0,
\end{equation}
   where one may define $\hat h \equiv \hat H^{-1} =0$ for $n=0$, consistent with the
   definitions in  (\ref{inv-crea-op}) and (\ref{inv-ann-op}).
   The Hamiltonian in this case is the
   inverse of the Hamiltonian of the standard (shifted) harmonic oscillator $ a^\dag a$.
   Unlike the  quantized harmonic oscillator,
   the spectrum of the new Hamiltonian system, $\hat h$,
   is not equally distanced (arises from the nonlinearity nature of the inverse system).

  \subsection{Introducing $|z, f\rangle^{(-1)}$
            as the coherent states associated to
            ${a^\dag}^{-1}$}

   Now, one may look for right eigenstate of the annihilation-like operator  $a_f$
   such that
 \begin{equation}\label{inv-css}
   a_f |z, f\rangle ^{(-1)} = z |z,f\rangle  ^{(-1)}.
 \end{equation}
   The superscript $(-1)$ on any state $|.\rangle$ (in the whole of the present paper)
   refers to the state corresponds to an "inverse" operator.
   A straightforward calculation shows that the state
   $|z, f\rangle^{(-1)}$ satisfies the eigenvalue equation (\ref{inv-css})
   has the following expansion in the Fock space
 \begin{equation}\label{nlcs-inv-bos}
   |z, f\rangle ^{(-1)}= N (|z|^2)^{-1/2}\sum_{n=0}^\infty  {\sqrt {n!}\; z^n }|n
   \rangle,
 \end{equation}
   with the normalization constant
 \begin{equation}\label{}
   N (|z|^2) = \sum_{n=0}^{\infty} n! \;|z|^{2n}
 \end{equation}
   which clearly diverges. Therefore, precisely speaking the
   eigenstate of the annihilation-like operator $a_f \equiv {a^\dag}^{-1}$ does really
   exist but  unfortunately it is physically
   meaningless (due to nonnormalizablity of the state).
   This is an expected result since the relation (\ref{diskff})
   determines the radius of convergence equal to 0 when $f(n)=\frac 1
   n$, i.e. for the case in hand. So, the states in (\ref{nlcs-inv-bos})
   can not actually  belong to the Hilbert space.

     \subsection{Introducing $|\tilde{z}, f\rangle^{(-1)}$
            as the dual family of $|z, f\rangle^{(-1)}$ }

    In what follows we will observe  that the  {\it dual
    family} of the states in (\ref{nlcs-inv-bos}) is well-defined.
    For this purpose, it is possible to  define
    two new operators $b_f$ and $b^\dag_f$ as   follows
\begin{equation}\label{}
  b_f = a \frac  {1}{f^\dag(\hat n)} = a \; \hat n,\qquad
  b^\dag_f =  \frac  {1}{f(\hat n)}a^\dag = \hat n\; a^\dag.
\end{equation}
   Thus, one has  $[a_f, \; b^\dag _f]=\hat I = [b_f, \; a^\dag _f]$.
   These properties  allow one to define the
   generalized (non-unitary) displacement operator as follows
\begin{equation}\label{}
  D_f(z)=\exp[zb^\dag _f -z^* a_f],
\end{equation}
   the action of which on the vacuum of the field gives the already
   obtained state in (\ref{nlcs-inv-bos}). But, according to the proposal
   has been recently  introduced in Refs. [11, 12]
   another  displacement operator may also be constructed as
 \begin{equation}\label{dis-dual}
    \widetilde D_f(z) = \exp[z a^\dag _f - z^* b_f],
 \end{equation}
  the action of which
  on the vacuum of the field gives
  a new set of nonlinear coherent states  as
 \begin{equation}\label{cs-invers-dual}
   |\widetilde z, f \rangle ^{(-1)}= \widetilde D_f(z) |0\rangle =
   \widetilde N (|z|^2)^{-1/2}
   \sum_{n=0}^{\infty} \frac {z^n} {(n!)^{3/2}}\;|n\rangle,
 \end{equation}
    where $z \in \C$. The normalization constant $\widetilde \N (|z|^2)$ can be
    obtained as
    \begin{equation}\label{}
    \widetilde N (|z|^2) = \sum_{n=0}^\infty
    \frac{|z|^{2n}}{(n!)^3} = {}_{0}F_2 ( 1, 1, |z|^2),
    \end{equation}
   where
    ${}_p F_q (a_1, \cdots, a_p; b_1, \cdots, b_p,; z)=
    \sum_{k=0}^\infty {\frac {(a_1)_k (a_2)_k
   \cdots (a_p)_k}{(b_1)_k (b_2)_k \cdots (b_q)_k}}
   \frac {z^k}{k!}$ is the
   generalized hypergeometric function and
   $(a)_m={\Gamma(a+m)}/{\Gamma(m)}$  with
   $\Gamma(m)$ the well-known Gamma function.
   It can be observed that these states can be
   defined on the whole space of complex plane.
   Nowadays the states in (9) and (29) are known under the name
   {\it "dual family"} or {\it"dual pair"} coherent states [11, 12].
   It can  be checked straightforwardly that the nonlinear coherent states
   in (\ref{cs-invers-dual}) are also the right eigenstates of  the deformed annihilation operator $b_f=a \hat n$.
   Thanks to J R Klauder {\it et al} for they
   established the resolution of the identity of the states in (\ref{cs-invers-dual})
   via the moment problem technique [13].

  \section{The inverse of the deformed annihilation (and creation) operator and
         the associated nonlinear coherent states}\label{sec-nl}
   Generalizing the proposed approach to
   the $F$-deformed rising and lowering operators
  \begin{equation}\label{}
     A= a F(\hat n), \qquad A^\dag = F^\dag (\hat n) a ^\dag,
  \end{equation}
   corresponding to nonlinear oscillator algebra, one can define
  \begin{equation}\label{inv-ann-nonL}
     A^{-1}= F^{-1}(\hat n) a^{-1}= \frac {1}{\hat n F(\hat n)} a^\dag
     \equiv \mathcal{F}^\dag (\hat n) a ^\dag,
   \end{equation}
   and
 \begin{equation}\label{inv-crea-nonL}
    {A^\dag}^{-1} = {a^\dag}^{-1} {F^\dag}^{-1} (\hat n)
    = a \frac {1}{ \hat n F^\dag(\hat n)} \equiv a \mathcal{F}(\hat n),
 \end{equation}
     where in the third steps of the derivation of the equations
     (\ref{inv-ann-nonL}) and (\ref{inv-crea-nonL}) the left equations of
     (\ref{af-1}) and (\ref{af-non}) have been used, respectively.
     In the continuation of the paper we shall call $F(n)$ as the
     {\it "original"} nonlinearity
     function.
     It is worth to mention two points. Firstly, is that the "generalized
     nonlinearity function" $\mathcal{F}(\hat n)$ has been defined in terms of the
     original nonlinearity function $F(n)$ as
 \begin{equation}\label{gen-nonlinearity}
     \mathcal{F}(\hat n) = \frac{1}{\hat n F^\dag(\hat n)},
 \end{equation}
     and secondly  since the original nonlinearity function $F(\hat n)$
     is  considered to be an operator valued function which
     generally can be complex [14], so is $\mathcal{F}(\hat n)$.
     The number states representations of the operators in
     (\ref{inv-ann-nonL}) and (\ref{inv-crea-nonL})
     take the forms:
 \begin{equation}\label{inv-ann-nonL-action2}
   A^{-1} \doteq  \sum_{n=0}^\infty   \frac {1}{\sqrt {n+1}F(n+1)}|n+1\rangle\langle n|.
 \end{equation}
 and  \begin{equation}\label{inv-crea-nonL-action1}
   {A^\dag}^{-1} \doteq \sum_{n=0}^\infty
    \frac {1}{\sqrt {n+1}F(n+1)}|n\rangle\langle n+1 |,
 \end{equation}
    It can be seen that
 \begin{eqnarray}\label{inv-crea-nonL-left-right}
    A A^{-1} &=&  {A^\dag}^{-1} {A^\dag}= \hat I,\\ \nonumber
    A^{-1} A &=& A^\dag {A^{\dag}}^{-1} = \hat I -| 0\rangle\langle  0 |,
 \end{eqnarray}
      which mean that $A^{-1}$ is the right inverse of $A$, and
      ${A^\dag}^{-1}$ is the left inverse of $A^\dag$, analogously to
      the interpretation of the inverse bosonic operators.
      With the help of the action of operators in  (\ref{inv-ann-nonL-action2})
      and (\ref{inv-crea-nonL-action1}) on the number states one has
 \begin{equation}\label{}
     {A^\dag}^{-1} | n \rangle = \frac{1}{\sqrt n F(n)} (1-\delta _{n, 0}) | n -1 \rangle,
 \end{equation}
    and
  \begin{equation}\label{}
    A^{-1} | n \rangle = \frac{1}{\sqrt {n+1} F(n+1)} | n +1 \rangle,
  \end{equation}
     where by definition ${A^{\dag}}^{-1}| 0 \rangle=0$.
     Therefore,  ${A^\dag}^{-1}$ and $A^{-1}$ act on the number states like annihilation and creation
     operators, respectively.
     We will rename reasonably  thus the generalized inverse operators in
    (\ref{inv-ann-nonL}) and (\ref{inv-crea-nonL})
    as
 \begin{equation}\label{}
        \mathcal{A}^\dag   \equiv  A^{-1} = \mathcal{F}^\dag (\hat n) a^\dag, \qquad
        \mathcal{A}  \equiv {A^\dag}^{-1} = a \mathcal{F}(\hat n),
 \end{equation}
    respectively.
   Note that the following commutation relation holds
 \begin{equation}\label{}
    [\mathcal{A},\; \mathcal{A}^\dag] =
    (\hat n+1)\left|\mathcal{F}(\hat n+1)\right|^2- \hat n \left|\mathcal{F}(\hat
    n)\right|^2
 \end{equation}
  which can be expressed  in terms of the $F$-function as
 \begin{equation}\label{}
   [\mathcal{A},\; \mathcal{A}^\dag] = \frac{1}{(\hat n+1)\left|F(\hat n+1)\right|^2}-
   \frac{1}{\hat n \left|F(\hat  n)\right|^2}, \qquad \mbox{for}\;\; n \neq
   0,
 \end{equation}
  and
 \begin{equation}\label{}
   [\mathcal{A},\; \mathcal{A}^\dag] = | 0 \rangle\langle 0 |, \qquad
   \mbox{for}\;\; n = 0.
 \end{equation}

    The dynamics of the "inverse nonlinear oscillator" may be described by the (inverse) Hamiltonian
 \begin{equation}\label{Hamilt-bob}
    \hat{\mathcal{H}} = {\mathcal{A}}^\dag \mathcal{A}
    = \frac {1} {\hat n | F(\hat n)|^2} \equiv \hat H ^{-1},\qquad \mbox {for}\;\; n\neq
    0\;,
 \end{equation}
    and $\hat{\mathcal{H}}=0$ for $n=0$.
    Interestingly, the Hamiltonian $\hat{\mathcal{H}}$ in (\ref{Hamilt-bob})
    is the inverse of the  Hamiltonian of the "original nonlinear oscillator" which is
    a familiar feature in the nonlinear coherent states context.

   Now, the  corresponding $F$-coherent states using the  algebraic definition
  \begin{equation}\label{}
   \mathcal{A} |z,\; \mathcal{F}\rangle ^{(-1)}  = z |z,\; \mathcal{F}\rangle  ^{(-1)},
  \end{equation}
   may be demanded.  A straightforward calculation shows that the states $|z,\; \mathcal{F}\rangle^{(-1)}$
   have the following expansions
 \begin{equation}\label{NLCS1}
   |z,\; \mathcal{F}\rangle  ^{(-1)}= \N (|z|^2)^{-1/2}
   \sum_{n=0}^{\infty} \frac {z^n} {\sqrt{n!} \;[\mathcal{F}(n)]!
   }\;|n\rangle.
 \end{equation}
   The states in (\ref{NLCS1}) when transformed in terms of the original nonlinearity
   function
   we started with ($F(n)$),
   take the following form
 \begin{equation}\label{NLCS2}
   |z,\; F\rangle  ^{(-1)} = \N (|z|^2)^{-1/2}
   \sum_{n=0}^{\infty}  {z^n\; \sqrt{n!} \; [F^\dag(n)]!}
   \;|n\rangle\;,
 \end{equation}
  where the definition
   $[\mathcal{F}(n)]! \doteq \frac{1}{n![F^\dag (n)]!}\;$ has been used and
   \begin{equation}\label{NLCS3}
      \N(|z|^2)= \sum_{n=0}^{\infty} n! \; |z|^{2n} ([F^\dag (n)]!)^2.
   \end{equation}
   With the particular choice of $F(n)=\frac 1 n$ in
   (\ref{NLCS2}) (or equivalently $\mathcal{F}(n) = 1$ in (\ref{NLCS1}))
   the standard coherent state in (3), known as {\it self-dual} states,   will be reobtained.

      Similar to the procedure led us to equation (\ref{res-fff}), the resolution of the identity requirement associated to the state in (\ref{NLCS1}) (or (\ref{NLCS2}))
      has been satisfied if a function $\eta(x)$ is found such
      that
 \begin{eqnarray}\label{res-f}
      \pi \int _0^{\RR} dx \eta (x) x^n &=&
      [n   {{\mathcal{F}}^\dag}^2 (n) ]!, \nonumber\\
      &=&  \left[\frac{1}{n {F^\dag}^2(n)}\right]!,
      \qquad n=0,1,2,\cdots ,
   \end{eqnarray}
    where $\RR$ is the radius of convergence determined by the disk
  \begin{equation}\label{disk}
    \D = \{z \in \C, \; |z|\leq \lim_{n\rightarrow \infty}
    [n {F^\dag }^2(n)]^{-1}\},
  \end{equation}
  centered at the origin in the complex plane.

  Related to the operators $\mathcal{A}^\dag$ and its conjugate
  $\mathcal{A}$, two conjugate operators can be defined as
 \begin{equation}\label{51}
   \mathcal{B}=a\frac{1}{\mathcal{F^\dag}(n)},\qquad \mathcal{B}^\dag=
   \frac{1}{\mathcal{F}(n)} a^\dag,
 \end{equation}
   such that the following canonical commutation relations hold
 \begin{equation}\label{52}
   [\mathcal{A}, \;\mathcal{B}^\dag]=\hat I = [\mathcal{B},\; \mathcal{A}^\dag].
 \end{equation}
  The relations in (\ref{51}) and (\ref{52}) enable one to define two generalized
  (non-unitary) displacement type operators
 \begin{equation}\label{D}
   D_{\mathcal{F}}(z) = \exp (z \mathcal{B}^\dag -z^* \mathcal{A}),
 \end{equation}
 and
 \begin{equation}\label{Dd}
   \widetilde D_{\mathcal{F}}(z) = \exp(z\mathcal{A}^\dag -z^* \mathcal{B}).
 \end{equation}
   By the action of  $D_{\mathcal{F}}(z)$ defined in (\ref{D}) on the fundamental state
   one readily finds that
 \begin{equation}\label{}
  D_{\mathcal{F}}(z) |0\rangle = |z,\; \mathcal{F}\rangle^{(-1)},
 \end{equation}
   which are exactly the states obtained in equations (\ref{NLCS1})
   and (\ref{NLCS2}) in terms of $\mathcal{F}(n)$ and
   $F(n)$, respectively. To this end, by the action of
   $\widetilde D_\mathcal{F}(z)$  in (\ref{Dd}) on the vacuum one gets a new set
   of states
 \begin{eqnarray}\label{DNLCS1}
    \widetilde D_\mathcal{F}(z) |0\rangle &=& |\widetilde{z},\;
    \mathcal{F}\rangle ^{(-1)}\nonumber\\
    &=& \widetilde \N(|z|^2)^{-1/2} \sum_{n=0}^\infty
    \frac{z^n}{\sqrt{n!}}\;[\mathcal{F}^\dag(n)]!\;|n\rangle.
 \end{eqnarray}
    The latter states  can be expressed in terms of the original
    function $F(n)$ as follows
 \begin{equation}\label{DNLCS2}
    |\widetilde{z},\;
    F\rangle ^{(-1)}= \widetilde \N(|z|^2)^{-1/2} \sum_{n=0}^\infty
    \frac{z^n}{\sqrt {n!} \; [n {F^\dag}(n)]!}\;|n\rangle,
  \end{equation}
  where
  \begin{equation}\label{DNLCS2-2}
    \widetilde \N(|z|^2) = \sum_{n=0}^\infty
    \frac {|z|^{2n}}{(n!)^3([F^\dag(n)]!)^2}.
  \end{equation}

      The resolution of the identity for
      the dual state in (\ref{DNLCS1}) (or (\ref{DNLCS2}))
      has been satisfied if a positive function $\widetilde \eta (x)$ is found such
      that
 \begin{eqnarray}\label{res-f}
      \pi \int _0^{\widetilde \RR} dx \widetilde \eta (x) x^n &=&
      \left[n   \frac{1}{{\mathcal{F}^\dag}^2 (n)} \right]!, \nonumber\\
      &=&  \left[n^3 {F^\dag}^2(n)\right]!,
      \qquad n=0,1,2,\cdots ,
   \end{eqnarray}
    where $\widetilde\RR$ is the radius of convergence determined by the disk
  \begin{equation}\label{disk}
    \widetilde{\D} = \{z \in \C, \; |z|\leq \lim_{n\rightarrow \infty}
    [n^3 {F^\dag }^2(n)]\},
  \end{equation}
  centered at the origin in the complex plane.

    Upon substituting $F(n)=1$ in (\ref{NLCS2}) and (\ref{DNLCS2})
    the states   in (\ref{nlcs-inv-bos})
    and (\ref{cs-invers-dual}) will be reobtained, respectively,  i.e. the
    dual family of coherent states associated to the inverse of bosonic operator.

   The states were introduced in (\ref{NLCS1}) and (\ref{DNLCS1})
   (or equivalently in (\ref{NLCS2}) and (\ref{DNLCS2}))
   are the dual pair (nonlinear)
   coherent states corresponding  to the generalized inverses of the
   deformed operators [11, 14].
   Comparing the state in (\ref{DNLCS2}) and the usual nonlinear
   coherent state in (9) shows that a multiplication factor $n!$
   appears in the denominator of the  expansion coefficient of the
   usual nonlinear coherent state.
   Notice that the existence of the factor $[F^\dag (n)]!$ in the
   expansion coefficients of (\ref{NLCS2}) and (\ref{NLCS3})
   (or (\ref{DNLCS2}) and (\ref{DNLCS2-2}))
   provides a good potentiality which allows one to use
   suitable nonlinearity functions $F(n)$ for constructing
   a wide variety of well-defined generalized coherent states associated  to inverse
   $F$-deformed operators.

   \section{Some physical realizations of the formalism and their nonclassical properties }
   Generally, a state is known as a nonclassical state (with no classical analogue) if
   the Glauber-Sudarshan $P(\alpha)$ function [15, 16] can not be interpreted
   as a probability density. However,
   in practice one can not directly apply this criterion to investigate
   the nonclassicality nature of a
   state [17]. So, this purpose has been frequently achieved
   by verifying {\it "squeezing, antibunching, sub-Poissonian
   statistics and oscillatory number distribution"}.
   A common feature of
   all the  above mentioned criteria is that  the corresponding
   $P$-function of a nonclassical state is not positive definite.
   Therefore, each of the above effects (squeezing or sub-Poissonian
   statistics which we will consider in the paper) is indeed sufficient
   for a state to possess nonclassicality signature.
 \begin{itemize}
   \item{\it Sub-Poissonian statistics}\\
   To examine the statistics of the states the  Mandel's $Q$-parameter is used,
   which characterizes the quantum states of light in the cavity.
   Mandel's $Q$-parameter has been defined as:
 \begin{equation}\label{Mandel}
    Q= \frac{\langle n^2 \rangle - \langle n \rangle ^2}{\langle n \rangle} -1.
 \end{equation}
    This quantity vanishes for {\it "standard coherent
    states"} (Poissonian), is positive  for {\it "classical"} (bunching effect),
    and negative
    for {\it "nonclassical"} light (antibunching effect).
 \item{\it Squeezing phenomena}\\
   Based on the following definitions of position and momentum operators
  \begin{equation}\label{}
     x= \frac {a+ a^\dag}{\sqrt 2}, \qquad  p= \frac {a- a^\dag}{\sqrt
     {2} \; i},
  \end{equation}
    the corresponding uncertainties will be defined as follows
  \begin{equation}\label{uncer}
    (\Delta x)^2 = \langle x^2\rangle - \langle x\rangle ^2,\qquad
    (\Delta p)^2 = \langle p^2\rangle - \langle p\rangle ^2.
  \end{equation}
   A state is squeezed in position or momentum quadrature if the uncertainty
   in the corresponding quadrature falls below the one's for the vacuum
   of the field; i.e. $(\Delta x)^2 < 0.5$ or $(\Delta p)^2 < 0.5$, respectively.
 \end{itemize}

   To give some physical realizations of the
   proposal, firstly one must
   specify the system with known "{\it nonlinearity function}" or
   "{\it discrete spectrum}" (these two quantities are related to each other through
   the relation
   $e_n=nf^2(n)$, where $e_n$ denotes the spectrum of physical system [9, 14]).
   At this stage in this section we shall concern with three particular systems:
   "hydrogen-like spectrum", "harmonious state" and "Gilmore-Perelomov
   representation of $SU(1,1)$ group",
   all of which the corresponding usual nonlinear coherent states and nonlinearity
   natures have been previously
   clarified.
   Squeezing effect and Mandel's $Q$-parameter for the obtained states
   in the paper  may be evaluated numerically.
   For this purpose one must calculate the expectation values
   expressed in  (\ref{Mandel}) and (\ref{uncer}) over any state of interest.
 \subsection{Hydrogen-like spectrum}
   As an important physical system we will accomplish in the present paper
   we want to apply our proposal onto the
   hydrogen-like spectrum. This quantum system is described by discrete spectrum:
 \begin{equation}\label{H-spect}
   e_n= 1- \frac  {1}{(n+1)^2}\; .
  \end{equation}
   The nonlinearity function in this case has been expressed as
   [9, 14]
  \begin{equation}\label{nonl-H-H}
   F_{H}(n)= \frac  {\sqrt{n+2}}{n+1}\; .
  \end{equation}
   The standard  nonlinear coherent state corresponding to
   this nonlinear function obtained with the help of
   (11) is restricted to a unit disk in
   the complex plane centered at the origin.
   In this subsection the nonclassicality nature of the dual pair
   of coherent states (according to the structural equations
   (\ref{NLCS2}) and (\ref{DNLCS2})) associated to hydrogen-like spectrum
   has been investigated.
   For the coherent states according to  (\ref{NLCS2}) in this example,
   the domain is restricted to $|z|< 1$, while  the domain would be
   $z \in \C$ when the states are constructed from   (\ref{DNLCS2}).
   The latter results are consistent with the  general feature occurs
   in the framework of the dual
   pair of coherent states,
   where if the domain of one set of a dual pair of coherent states
   is the whole of the complex plane, that
   of the other set would be the unit disk and vice versa  \cite{Roknizadeh2004}.
   Anyway, for instance, to verify the resolution of the identity for the corresponding dual
   states according to (\ref{DNLCS2}) we use the definition of
   Meijer's $G$-function together with the inverse Mellin theorem \cite{Mathai
   1973}:
 \begin{eqnarray}\label{Meiger}
    \int_0^\infty dx x^{s-1} G_{p, q}^{m, n}
     \left( \alpha x \Big|
   \begin{array}{cccccc}
      a_1, & \cdots, & a_n, & a_{n+1}, & \cdots, & a_p  \\
      b_1, & \cdots, & b_m, & b_{m+1}, &\cdots, & b_q
   \end{array}
     \right)   =   \frac{1}{\alpha ^s} \nonumber\\ \times \frac  {\Pi_{j=1}^{m}\Gamma
     (b_j +s) \Pi_{j=1}^{n}\Gamma (1-a_j-s)} {\Pi_{j=n+1}^{p}\Gamma (a_j+s)
     \Pi_{j=m+1}^{q}\Gamma (1-b_j-s) }.
 \end{eqnarray}
   So for instance, the function $\widetilde{\eta}(x)$ satisfying equation
   (\ref{res-f}) with nonlinearity function introduced
   in (\ref{nonl-H-H})  may be given in terms of the
   Meijer's $G$-function by the expression
 \begin{eqnarray}\label{Meiger-h}
   \widetilde{\eta}(x) = G^{4, 0}_{2, 0}
     \left(  x \Big|
   \begin{array}{cc}
      0, 0, 0, 2 &  .  \\
         0, 0 & .
   \end{array}
      \right).
      \end{eqnarray}
    Thus, the associated weight function satisfies the resolution
    of the identity for these set of states can be calculated as
    $\widetilde{\W}(x)= \widetilde{\eta}(x) (1+x)\;  _{0}F_{1}(3,
    x)$, where $_{0}F_{1}$ is the regularized confluent hypergeometric function
    and $\widetilde{\eta}(x)$ is determined in (\ref{Meiger-h}).
    Now, the numerical results for the dual pair of coherent states
    according to  (47) and (57) with the nonlinearity function in (65)
    will be displayed in figures (1)-(4).
    Although the dual pair of coherent states are defined on different domains,
    for the sake of comparison
    our numerical calculations for both of them  presented just for $|z|< 1$.
  In figure (1) the uncertainties in $x$ and $p$ have been
  plotted in the respected domain as a function of
  $z \in \R$ utilizing  (\ref{NLCS2}). The squeezing in
  $p$-quadrature has been shown  for real
  $z < 1$.
  Figure (2) is the same as figure (1)
  except that the equation (\ref{DNLCS2}) has been considered. The squeezing in
  $x$-quadrature has been occurred for real
  $z$.
   Our further computations   when $z >> 1$ for the example in hand upon using
   (57), has defined on the whole complex plane, indicated that
   the variances in $x$ (solid line) and $p$ (dashed line) tend  respectively at about
   $\cong 0.25$ and  $\cong 1$. So, the squeezing in
   $x$ is visible for any real value of $z$.
  The three-dimensional graph of Mandel's $Q$-parameter as a function of
  $z \in \C$ has been shown in figure (3)
  for the states constructed according to  (47). As it can be seen the
  sub-Poissonian statistics is restricted to a finite range of values of $z$
  near  $z < 1$.
  Figure (4) is the same as figure (3) when  (\ref{DNLCS2}) is
  used. In  this case the sub-Poissonian exhibition has been
  occurred for all values of $z \in \C$.
  So, in view of this result, the latter are fully nonclassical states,
  in the sense that they have nonclassical nature within the whole
  permitted range of $z$ values.
 \subsection{Harmonious states}
  Harmonious states characterized by the nonlinearity function
 \begin{equation}\label{nonl-H-Hs}
    F_{HS}(n)=\frac{1}{\sqrt n}\;
 \end{equation}
    is in considerable attention in quantum optics.
   It can be observed that the
   lowering operator constructed from the nonlinearity function in
   (\ref{nonl-H-Hs}) is equivalently
   the nonunitary  Susskind-Glogower operator
   $\exp (i \hat{\Phi}) = a (a^{\dag} a)^{-1/2}$ \cite{Agarwal2}.
   It has been shown that the probability operator measures
   generated by the latter operator yields
   the maximum likelihood quantum phase
   estimation for an arbitrary input state  \cite{Shapiro}.
    Inserting $F_{HS}(n)$ from (68) in  (\ref{NLCS2})
    one will reobtain the harmonious
    states introduced and discussed in detail by Sudarshan [21], restricted again to a
    unit disk in the complex plane.
    On the other hand, substituting this nonlinearity function into  (\ref{DNLCS2})
    yields the following generalized coherent states:
   \begin{equation}\label{dual-harmonous}
    |\widetilde{z},\;
    F\rangle ^{(-1)} = \widetilde \N(|z|^2)^{-1/2} \sum_{n=0}^\infty
    \frac{z^n}{n!}|n\rangle,
   \end{equation}
    with the normalization constant
  \begin{equation}\label{}
     \widetilde \N(|z|^2) = \sum_{n=0}^\infty
     \frac {|z|^{2n}}{(n!)^2}=I_0(2\sqrt{|z|^2}),
  \end{equation}
  where in the last expression $I_0(x)$ is the
  modified Bessel function of the first kind. The domain of this
  set of coherent states is the whole complex plane.
  The resolution of the identity in this case is satisfied with the
  choice of a density function may be determined as $\widetilde \eta (x) =
  2 K_0 (2\sqrt x)$.
  So the associated weight function will be $\widetilde \W(x) =
  2 K_0 (2\sqrt x) I_0(2\sqrt x)$,
  where $I_0$ and $K_0$ are the modified Bessel functions of
  the first and third kind, respectively  \cite{kps}.

  We refrain from graphical representations, since the numerical results
  with the nonlinearity function (68) are closely the same as the
  hydrogen-like states have been illustrated
  in subsection 4.1. This  fact may be expected,
  because the two nonlinearity functions in (\ref{nonl-H-H})
  and (\ref{nonl-H-Hs}) are not far from each other especially for large $n$.

  \subsection{Gilmore-Perelomov  representation of $SU(1, 1)$ group}
   As final example we are interested in is  the
   Gilmore-Perelomov (GP) coherent state of $SU(1, 1 )$ group
   whose the number state representation read as \cite{Perelomov}:
 \begin{equation}\label{su}
   |z, \kappa\rangle^{su(1,1)}_{GP} = N(|z|^2)^{-1/2}\sum_{n=0}^\infty
   \sqrt{\frac{\Gamma(n+2\kappa)}{n!}}z^n \; |n\rangle,
 \end{equation}
    where $N(|z|^2)$  is the normalization constant which can
    be written in closed form as follows
  \begin{equation}\label{}
    N(|z|^2)= (1-|z|^2)^ {-2 \kappa} \Gamma(2 \kappa).
  \end{equation}
     Notice that the expansion in  (\ref{su}) the label
    $\kappa$ takes the discrete values $1/2, 1, 3/2, 2, 5/2,
    \cdots$.
    The  nonlinearity function in this case is determined as [9]
  \begin{equation}\label{fsu}
    F(n, \kappa)=\frac{1}{\sqrt{n+2\kappa -1}}.
  \end{equation}
  According to equation (\ref{NLCS2}) the corresponding
  coherent state associated to inverse $F$-deformed
  annihilation-like operator using $F(n, \kappa)$ in (73) takes the
  following decomposition in number states:
 \begin{equation}\label{su-inv}
   |z, F\rangle^{(-1)}_{GP} = \N(|z|^2)^{-1/2}\sum_{n=0}^\infty
   \sqrt{\frac{n!}{\Gamma(n+2\kappa-1)}}\;z^n \; |n\rangle,
 \end{equation}
     with the normalization constant
  \begin{equation}\label{norm}
    \N(|z|^2)= \frac{ {}_2 F_{1} (\{1,1\}; \{2 \kappa\}; |z|^2)} {\Gamma(2
    \kappa)},
  \end{equation}
   where  ${}_p F_{q}(\overrightarrow{a}; \overrightarrow{b}; x)=
   {}_p F_{q} (\{a_1, \cdots, a_p\}; \{b_1, \cdots, b_q \}; x)$ is the
   generalized hypergeometric function.
   These states can be defined in the unit disk.
   Similarly, the explicit expansion of the dual
   family of (\ref{su-inv}) can be obtained easily by inserting
   $F(n, \kappa)$ from (\ref{fsu}) into the structural equation (\ref{DNLCS2})
   which has the whole complex plane domain.
   To investigate
   the resolution of the identity associated to the
   states have been constructed according to the structural equation (\ref{DNLCS2})
   one gets
 \begin{eqnarray}\label{Meiger-su}
  \widetilde{\eta}(x) = G^{3, 0}_{1, 0}
     \left(  x \Big|
   \begin{array}{cc}
      .& 2 \kappa -1  \\
        0, 0, 0   & .
   \end{array}
      \right),
      \end{eqnarray}
      where again we have used (\ref{Meiger}).
      Therefore, the corresponding  weight function satisfies
      the resolution of the identity in this case can be evaluated as
      $\widetilde{\W}(x)=\Gamma(2 \kappa) \widetilde{\eta}(x) \; _{1}F_{1}(2\kappa; 1;
      x)$,
      where $  _{1}F_{1}$ is the Kummer confluent hypergeometric function
      and $\widetilde{\eta}(x)$ has been  defined in equation (76).
   Now, we discuss the numerical results of $SU(1, 1)$ group in  figures (5)-(10).
   In figure (5) the uncertainties in $x$ and $p$ have been shown with respect to
   $z  \in \R $ for different values of $\kappa$ according to the construction of states
   using  (\ref{NLCS2}).
   The squeezing in
   $p$-quadrature is visible for all values of $z < 1 $,
   irrespective of $\kappa$ values. It is seen that the maximal
   squeezing occurred for $\kappa=\frac  1 2$.
   Figure (6) is the same as figure (5)
   when  (\ref{DNLCS2}) has been considered. In this case
   squeezing has been shown in $x$-quadrature for all values of $\kappa$, when $z\in \R$.
   The three-dimensional graphic representation of Mandel's
   $Q$-parameter for the state corresponding to (\ref{NLCS2})
   is plotted in figure (7).
   The sub-Poissonian statistics in a finite range of values of $z$ has been
   shown (when both the real and imaginary parts of  $z$ are near $1$,
   the upper bound of $z$).
   Figure (8) is a typical two-dimensional plot of figure (7) when the real part of $z$
   is fixed at particular value, i.e. $x=0.8$.
   This figure may be useful to illustrate figure (7) in detail.
   Figure (9) is the same as figure (7) when (\ref{DNLCS2}) is
   used. The sub-Poissonian statistics has been occurred for all values of $z \in
   \C$.
    According to our calculations for different values of $\kappa$, the
    negativity of the Mandel's $Q$-parameter decreases
    (for the states of the type (\ref{DNLCS2})) with
    increasing $\kappa$ (figure (10) shows this fact when compares with figure (9)).
    But the sub-Poissonian nature of the latter states in the complex plane
    preserves for all allowed values of $\kappa$.
    Our further calculations show that by
    increasing the real and imaginary parts of $z$
    the Mandel's $Q$-parameter fixes
    at a certain negative value between $0$ and $-1$.
    To this end, as stated in the case of Hydrogen-like spectrum,
    the latter states of $SU(1, 1)$ constructed utilizing
    equation (57) are also fully nonclassical states.

 \section{Summary and conclusion}

    The large number of applications of coherent states in various
    areas of physics motivates to enlarge them,
    so looking for novel definitions
    and new classes of states is of much interest.
    In this paper, based on the {\it "nonlinear coherent states method"},
    a formalism for the construction of coherent state associated
    to the {\it "inverse (bosonic) annihilation-like operator"}
    has been introduced.
    Although the latter was ill-defined,
    their dual family has been obtained in a proper fashion.
    Generalizing the concept, the
    {\it "inverse nonlinear ($F$-deformed) ladder operators"} corresponding to the
    deformed rising and lowering operators involved in the
    nonlinear coherent states of
    quantum optics, and {\it "the associated  nonlinear ($F$-deformed)
    coherent states"} have been
    introduced.
    The presented formalism provides a framework that
    by virtue of the generalized coherent states have been
    previously  introduced (with known nonlinearity functions or
    corresponding to any exactly solvable potential with discrete spectrum)
    \cite{Sunilkumar, Shreecharan}
    it will be possible to construct new classes of generalized coherent states
    associated to generalized
    inverse  ($F$-deformed annihilation-like) operators.
    So, a large set of generalized coherent states in addition to their dual
    families can be constructed in the field of quantum optics.
    We hope that the introduced states may find
    their useful applications in different physical situations,
    both theoretically and experimentally.

 \vspace {.5 cm}
   {\bf Acknowledgement:}
   {The author would like
   to acknowledge Prof.  R. Roknizadeh for introducing him the quantum
   optics field of research
   at first  and  the coherent states as the venerable objects in this domain
   and from the referees for their valuable comments  which improve the paper appreciably.
   Also thanks to the research council  of
   Yazd University for their financial supports of this project.}
\vspace {3 cm}
\newpage
{\bf FIGURE CAPTIONS}
  \vspace {1.5 cm}

  {\bf FIG. 1} The uncertainties  in $x$ (the solid curve)
     and $p$ (the dashed curve) as a function of
  $z\in \R$ for hydrogen-like spectrum according to the equation (47).

  \vspace {.5 cm}

  {\bf FIG. 2} The same as figure (1)
   except that the equation  (57) has been considered.

  \vspace {.5 cm}

  {\bf FIG. 3.} The graph of Mandel's $Q$-parameter as a function of
  $z \in \C$, for the  hydrogen-like spectrum according to the equation  (47).

  \vspace {.5 cm}

  {\bf FIG. 4} The same as figure (3) except that the equation  (57) has been considered.

  \vspace {.5 cm}

  {\bf FIG. 5}
   The uncertainties in $x$ and $p$ versus $z \in \R$ for the GP
   representations of $SU(1, 1)$ group according to the equation (47)
   for different values of $\kappa$; $\kappa=0.5$ (solid lines), $\kappa=1$ (dashed lines)
   and  $\kappa=1.5$ (dotted lines). Squeezing in $p$ is observed in all
   cases.

  \vspace {.5 cm}

  {\bf FIG. 6} The same as figure (5)
   except that the equation  (57) has been considered.
   Squeezing in $x$ is observed in all cases.

  \vspace {.5 cm}

  {\bf FIG. 7} The Mandel's $Q$-parameter
  as a function of  $z \in \C$ for GP representation of $SU(1, 1)$ group,
  according to the  equation (47), ($\kappa$ is set equal to $1/2$).

  \vspace {.5 cm}

  {\bf FIG. 8} The two-dimensional Mandel's $Q$-parameter of figure (7)
   as a function of  $y$ (when $x= 0.8$) for GP representation of $SU(1, 1)$ group,
   according to the  equation (47), ($\kappa$ is set equal to $1/2$).

  \vspace {.5 cm}

  {\bf FIG. 9} The same as figure (7)
    except that the equation (57) has been considered with $\kappa=1$.

  \vspace {.5 cm}

  {\bf FIG. 10}  The same as figure (7)
    except that the equation  (57) has been considered with $\kappa=3$.



 \include{thebibliography}


 \end{document}